# Title: Valley Phonons and Exciton Complexes in a Monolayer Semiconductor

**Authors:** Minhao He[1†], Pasqual Rivera[1†], Dinh Van Tuan[2], Nathan P. Wilson[1], Min Yang[2], Takashi Taniguchi[3], Kenji Watanabe[3], Jiaqiang Yan[4,5], David G. Mandrus[4-6], Hongyi Yu[7], Hanan Dery[2,8#], Wang Yao[7#], & Xiaodong Xu[1,9#]

**Affiliations:**
[1] Department of Physics, University of Washington, Seattle, Washington 98195, US A
[2] Department of Electrical and Computer Engineering, University of Rochester, Rochester, New York, 14627, USA
[3] National Institute for Materials Science, Tsukuba, Ibaraki 305-0044, Japan
[4] Materials Science and Technology Division, Oak Ridge National Laboratory, Oak Ridge, Tennessee, 37831, USA
[5] Department of Materials Science and Engineering, University of Tennessee, Knoxville, Tennessee, 37996, USA
[6] Department of Physics and Astronomy, University of Tennessee, Knoxville, Tennessee, 37996, USA
[7] Department of Physics and Center of Theoretical and Computational Physics, University of Hong Kong, Hong Kong, China
[8] Department of Physics and Astronomy, University of Rochester, Rochester, New York 14627, USA
[9] Department of Materials Science and Engineering, University of Washington, Seattle, Washington, 98195, USA

[†] These authors contributed equally to the work.
[#]Correspondence to xuxd@uw.edu; wangyao@hku.hk; hanan.dery@rochester.edu

**Abstract:** The coupling between spin, charge, and lattice degrees of freedom plays an important role in a wide range of fundamental phenomena. Monolayer semiconducting transitional metal dichalcogenides have emerged as an outstanding platform for studying these coupling effects because they possess unique spin-valley locking physics for hosting rich excitonic species and the reduced screening for strong Coulomb interactions. Here, we report the observation of multiple valley phonons – phonons with momentum vectors pointing to the corners of the hexagonal Brillouin zone – and the resulting exciton complexes in the monolayer semiconductor $WSe_2$. From Landé g-factor and polarization analyses of photoluminescence peaks, we find that these valley phonons lead to efficient intervalley scattering of quasi particles in both exciton formation and relaxation. This leads to a series of photoluminescence peaks as valley phonon replicas of dark trions. Using identified valley phonons, we also uncovered an intervalley exciton near charge neutrality, and extract its short-range electron-hole exchange interaction to be about 10 meV. Our work not only identifies a number of previously unknown 2D excitonic species, but also shows that monolayer $WSe_2$ is a prime candidate for studying interactions between spin, pseudospin, and zone-edge phonons.

**Maintext:**

Electron-phonon interaction is a ubiquitous process in solids. In monolayer semiconducting transition metal dichalcogenides (TMDs), the broken inversion symmetry and strong spin-orbit coupling leads to the well-known spin-valley coupling of band edge electrons[1-6]. The emergent valley-contrasting properties not only impact quasiparticles, but also are expected to give rise to new physics involving zone edge phonons, or valley phonons, which are collective lattice oscillations at the corners of hexagonal Brillouin zone (±$K$ points). These phonons have been predicted to play an important role in spin and valley pseudospin relaxation through phonon-assisted intervalley scattering[7,8]. Additionally, valley phonons can possess chirality with intrinsic pseudo-angular momentum[9], which has recently attracted wide attention[10-13]. Such chiral phonons have nontrivial Berry curvature and are predicted to give rise to valley phonon Hall effect, a counterpart of valley Hall effect of electrons in 2D semiconductors[14-16]. Despite the importance of valley phonons, experimental progress in understanding their properties has been limited, since it is challenging to probe phonons with large momentum vectors.

Here, we identify the signatures of multiple valley phonons in monolayer $WSe_2$. The monolayer $WSe_2$ hosts stable and long-lived dark exciton and trion as ground states[17-22]. The resulting accumulated exciton and trion populations are highly desirable for studying their interactions with phonons. We found that three valley phonons facilitate efficient spin-conserving intervalley scattering, which results in a series of dark exciton and dark trion phonon-replicas in the low temperature photoluminescence (PL) spectrum. The sign and magnitude of Landé effective g-factors of the various replicas, together with their PL helicity under optical pumping, reveals that the spin-preserving intervalley scattering of the electron is more efficient than its intravalley spin-flip during the dark exciton/trion formation process. This results in a surprising finding: the single electron in both the positive dark trion and intervalley exciton resides in the valley opposite to that which the optical pump is coupled to. Moreover, the identified intervalley exciton resonance enables us to infer a short-range electron-hole exchange interaction of ~10 meV by extracting the energy splitting between intervalley and dark (intravalley) excitons.

The samples are exfoliated monolayer $WSe_2$ encapsulated between thin flakes of hexagonal boron nitride (hBN). Few-layered graphene serves as a local bottom gate for electrostatic control of the monolayer carrier density (Methods). Figures 1a and 1b are an optical microscope image and schematic of a representative device, respectively. Figure 1c shows the PL intensity plot as a function of gate voltage (V) and photon energy, at a temperature of 1.6 K. The laser energy is 1.775 eV with right circularly polarized ($\sigma^+$) excitation and unpolarized detection. The full helicity-resolved gate dependent spectra are shown in Fig. S1. Monolayer WSe2 hosts a rich spectrum of excitonic species[23-28]. Several previously identified excitonic states are indicated in the figure, including the neutral bright exciton ($X^0$), bright trions ($X^\pm$)[29-33], the intravalley spin-forbidden dark exciton ($D^0$)[17-20,34], and dark trions ($D^\pm$)[18,35]. The recently identified zone-center $\Gamma_5$ – or $E''$ – phonon replicas below both neutral ($D^0_{\Gamma_5}$)[12,13] and charged dark excitons ($D^\pm_{\Gamma_5}$)[12] are also resolved, as indicated by black arrows. The focus of this paper is to understand the valley phonon origin of several previously unidentified PL peaks. In particular, we focus on the triplet and quadruplet PL peaks outlined by dashed black boxes, and those pointed at by the white arrows in Fig. 1c. We have measured multiple samples that exhibit similar spectrum (Fig. S2), and the nearly identical power dependence of the peaks of interest across several samples rules out the possibility of them arising due to defect states (Fig. S3).

We first consider the quadruplet PL peaks in the hole doping regime. Fig. 2a shows the circular polarization resolved PL with $\sigma^+$ polarized excitation. The photon energy is relative to the positive dark trion ($D^+$), and the intensity of the quadruplets is multiplied by a factor of 4 to emphasize these weak spectral features. We ascribe the four peaks to bright replicas of the dark trion, mediated by interactions with phonons. The quadruplets are labeled as $D^+_{K_3}$, $D^+_{\Gamma_5}$, $D^+_{K_1}$, $D^+_{K_2}$, which correspond to their origin from coupling of $D^+$ with valley phonons $K_3$, $K_1$, and $K_2$, and the zone-center phonon $\Gamma_5$. This nomenclature has its origin in the Koster notation of the $K$-point irreducible representations of the $C_{3h}$ point double-group corresponding to the symmetry of the monolayer semiconductors (Table S3, character table).

The valley phonon replicas $D^+_{K_3}$, $D^+_{K_1}$, $D^+_{K_2}$ are 26, 18, and 13 meV below $D^+$, respectively. The association with zone-edge phonons $K_3$, $K_1$, and $K_2$ is two-fold. The first reason is that the energy differences between $D^+$ and its replicas match the energies of the associated phonon modes (Fig. S5, phonon spectrum). The second reason is rooted in the selection rules of electron-phonon coupling[7]; intervalley transitions of electrons and holes are mediated only by specific phonons (see discussions in SI-4 and SI-5). Finally, we note that $D^+_{\Gamma_5}$ appears 22 meV below $D^+$, which is consistent with the recently reported $\Gamma_5$ phonon replica of $D^+$.[12]

To further examine these assignments, we performed magneto-PL to extract the effective Landé g-factors of the various states, which are then used to identify the spin and valley indices of the constituent electrons and holes in the excitonic states[36-40]. The extracted g-factors of the states of interest are listed in Table 1. The g-factors of other states are listed in Table S1, and a detailed analysis can be found in SI-3. Briefly, for intravalley electron-hole recombination, the magnitude of the g-factor is about 4 for parallel electron and hole spins[36,37,39,40], and about 9 when the two spins are anti-parallel[12,13,21]. Meanwhile, if the recombination involves electron and hole from opposite valleys but with parallel spins, a value about 13 is expected.

Figure 2b shows the circular polarization resolved PL intensity as a function of out-of-plane magnetic field. The energy scale is relative to the position of $D^+$ at zero applied magnetic field. Both the cross pattern of $D^+$ in Fig. 2b and its unpolarized light emission (see Fig. 2a) result from the underlying out-of-plane dipole orientation[12], and are hallmarks of direct intravalley recombination of dark trions. Following the convention of valley Zeeman splitting as $\Delta = E(\sigma^+) - E(\sigma^-)$, where $E(\sigma^+)$ and $E(\sigma^-)$ are the peak energies of the $\sigma^+$ and $\sigma^-$ polarized PL components, we obtained $g(D^+) = -8.6$. The obtained g-factor of $D^+$ is therefore consistent with expectations for the direct recombination through the intravalley spin-flip transition.

The Zeeman shifts of the quadruplet PL peaks underpin their origin as phonon replicas of the dark positive trion $D^+$. The extracted g-factors of -13.4, 12.2, and -13.0 for $D^+_{K_3}$, $D^+_{K_1}$, $D^+_{K_2}$, respectively, are consistent with intervalley recombination of electron and hole. Note that the g factor sign of $D^+_{K_1}$ is opposite to others, which will be discussed later. While the intervalley recombination is naturally forbidden for delocalized exciton complexes, because of the large momentum mismatch, emission of a valley phonon can supply the required momentum, resulting in phonon-assisted luminescence of the otherwise dark states. Combined with the concurrence of similar valley phonon energies and energy difference between the peaks and $D^+$, we conclude that $D^+_{K_3}, D^+_{K_1}, D^+_{K_2}$ are $K_3$, $K_1$, and $K_2$ valley phonon replicas of $D^+$.

The polarization of the $D^+_{\Gamma_5}$ peak highlights the importance of valley phonons in the formation process of $D^+$. From Fig. 2a, we observe that $D^+_{\Gamma_5}$ is cross-circularly polarized. The valley optical selection rules dictate that $\sigma^+$ excitation creates an electron and hole in the $+K$ valley, while $\sigma^-$ polarized emission can only happen through spin-conserved electron-hole recombination in the $-K$ valley. In addition, $g(D^+_{\Gamma_5}) = $ -9.7 indicates the intravalley electron-hole recombination nature of the peak. The cross-polarized emission of $D^+_{\Gamma_5}$ therefore leads to the surprising conclusion that $\sigma^+$ excitation results in $D^+$ with the single electron located in the -K valley: i.e. $D^+$(-K).

The cross polarization of $D^+$ can be understood by considering the impact of phonons on the electron relaxation pathways following photoexcitation. The $D^+$ population is created by optical pumping of the spin-conserved interband transition (Fig. 2c), followed by the relaxation of the electron from the higher energy spin-valley locked sub-band to the lower energy one. The latter step requires either a spin-flip or a valley-flip. From symmetry analysis, the $\Gamma_5$ phonon can lead to the intravalley spin-flip relaxation of electron from upper to lower conduction band[7], but it cannot cause intervalley scattering. On the other hand, spin-conserving intervalley scattering of the electron with $K_3$ phonon is a symmetry-allowed zeroth-order channel[8]. Fig. 2c illustrates the intervalley electron-phonon relaxation process that leads to the formation of $D^+(-K)$. First, $\sigma^+$ excitation creates electron in the spin up conduction band in the $K$ valley. Assisted by the $K_3$ valley phonon (Fig. 2d), this spin up electron is then scattered into the spin up band in the $-K$ valley, forming $D^+(-K)$ with two holes separately located at the top of $\pm K$ valleys. $D^+(-K)$ then couples to $\sigma^-$ polarized photon by emitting $\Gamma_5$ phonon, as shown in Fig. 2e. Evidently, the observed cross-polarized $D^+_{\Gamma_5}$ emission implies that the valley-flip rate exceeds the spin-flip one in the relaxation of electron.

Having established that $\sigma^+$ polarized excitation results in $D^+(-K)$, the understanding of both $D^+_{K_3}$ and $D^+_{K_2}$ is straightforward. As indicated in Fig. 2f, the spin up electron in the lower $-K$ sub-band is virtually scattered to the higher $+K$ sub-band by emitting either a $K_3$ or $K_2$ valley phonon(see SI-6 for further analysis of $K_2$). The spin-conserving intervalley scatter then allows for recombination with the hole in the $+K$ valley, emitting $\sigma^+$ polarized photon with energy either 26 meV ($K_3$) or 13 meV ($K_2$) below $D^+$. The g-factors of $D^+_{K_3}$ and $D^+_{K_2}$ are nearly equal, about -13.4 and -13.0, respectively, and correspond to expected values for the intervalley spin-conserving electron-hole recombination.

Moreover, the measured amplitude of $D^+_{K_3}$, which is several times stronger than that of $D^+_{K_2}$, is also consistent with group-theory selection rules. In particular, the selection rules dictate that, in pristine monolayer WSe$_2$, only the $K_3$ phonon mode can induce intervalley electron transitions between conduction band edges at the high symmetry $K$ and $-K$ points[7]. Meanwhile, intervalley electron transitions that are mediated by other $K$-point phonon modes are higher-order processes with correspondingly smaller amplitudes that involve electron states in the neighborhood of $\pm K$. Such relatively weak processes can be amplified by several possible sources such as localization next to defects, breaking WSe$_2$ mirror inversion symmetry by hBN encapsulation, or WSe$_2$/hBN moire superlattice providing in plane momentum.

The positive g-factor of $D_{K_1}^+$ is a signature of the interaction of the hole and $K_1$ valley phonon. According to the group-theory selection rules, $K_1$ is the only phonon mode that enables the spin-conserving intervalley transition between valence-band states at $K$ and $-K$ points (i.e., it is a zeroth-order process)[7]. As shown in Fig. 2g, the hole in the higher valence band of $K$ valley is virtually scattered to the lower valence band of the $-K$ valley with the same spin orientation via emission of a $K_1$ valley phonon, forming an intermediate virtual B trion. The recombination of the electron with this scattered hole in the $-K$ valley results in $D_{K_1}^+$, with energy 18 meV below $D^+$. While the initial and final states have the same spin-valley configuration as in $D_{K_3}^+$ and $D_{K_2}^+$, the emission from $D_{K_1}^+$ is $\sigma^-$ polarized. Therefore, the g-factor of $D_{K_1}^+$ is expected to have similar magnitude, but opposite sign, compared to the g-factors of $D_{K_3}^+$ and $D_{K_2}^+$, as we have observed. Note that although the interaction between the hole and valley phonon $K_1$ is relatively strong, the coupling of $D^+(-K)$ (or A trion) with the virtual B trion state is ~400 meV detuned, which is much larger than in the case of $K_3$ intervalley electron scattering. Therefore, the PL intensity of $D_{K_1}^+$ is expected to be several times weaker than that of $D_{K_3}^+$, which is in agreement with our observation (see SI-5).

The valley phonon replicas of the negative dark trion $D^-$ can be understood using similar analysis as above. Fig. 3a shows the helicity-resolved PL under $\sigma^+$ polarized excitation. The lower energy spectral features, denoted as $T_1$, $D_{K_3}^-$, and $D_{\Gamma_5}^-$ show appreciable co-circular polarization. Fig. 3b shows the PL intensity as a function of magnetic field and photon energy, with $\sigma^-/\sigma^-$ polarized excitation/detection (the $\sigma^+/\sigma^+$ results are shown in Fig. S7). The extracted g-factor of $D_{K_3}^-$ is -12.5, indicating intervalley recombination, and the energy is 26 meV below $D^-$. As such, we can identify that $D_{K_3}^-$ originates from the interaction of $-K$ valley electron with $K_3$ phonon, as depicted in the inset of Fig. 3a. The g-factor of $D_{\Gamma_5}^-$ (-9.9) is nearly the same as that of $D^-$ (-9.5), and as expected, the peak appears 22 meV below $D^-$.

The observation of triplet peak pattern in the PL spectrum under electron doping is slightly different from the quadruplet one observed under hole doping. Since $T_1$ is more intense than $D^-$, and its g-factor is about -4.5, it is not likely to be a phonon replica of $D^-$. Its origin is unknown, but $T_1$ is near the spectral range of the $K_2$ and $K_1$ valley phonon replicas, and thus obscures them entirely. Nevertheless, the large difference in g-factor of $T_1$ from that of $D^-$ allows us to identify $D_{K_2}^-$ in magneto-PL, since the states shift away from one another. We note that there is a faint line in Fig. 3b, appearing on the high energy shoulder of $T_1$ at high magnetic field. We ascribe this peak to $D_{K_2}^-$ based on its g-factor of about -13.6, and its energy difference from $D^-$ of 13 meV, both of which are in good agreement with expectations for $K_2$ phonon replica.

Meanwhile, $D_{K_1}^-$ becomes evident when the excitation laser is resonant with the bright neutral exciton ($X^0$). Fig. 3c presents the PL intensity as a function of gate voltage and photon energy, with $\sigma^+$ polarized excitation laser in resonance with $X^0$ (1.733 eV) and $\sigma^-$ polarized detection. For V = 0.2V, arrow on side of Fig. 3c, $D_{K_1}^-$ becomes apparent at 18 meV below $D^-$. Fig. 3d shows magneto-PL with the same cross-polarized polarization ($\sigma^+/\sigma^-$) (see Fig. S8 for complete data set). We find that $D_{K_1}^-$ has a positive g-factor of 12.2, which is identical to that of $D_{K_1}^+$. Therefore, despite the initial appearance, the four phonon replica states observed under hole-doping also appear under electron-doping conditions. Moreover, we observe similar magnitude and sign of g-factors, as well as energy separation from the parent dark trion state.

The valley phonon-assisted momentum relaxation mechanism also produces phonon replicas for neutral dark excitons. Fig. 4a shows the helicity-resolved PL spectra near charge neutrality (V = -0.2V). The energy axis is offset from $I^0$, which is the sharp line 32 meV below the bright neutral exciton in Fig. 1c. We also observe strong direct recombination from the neutral intravalley dark exciton $D^0$, at 42 meV below $X^0$. We confirm that $D^0$ has negligible circular polarization, resulting from its out-of-plane dipole orientation, and the extracted g-factor (-9.1) is in good agreement with previously reported values[12,13,21]. Magneto-PL measurements with linear excitation (V) and σ+/σ- collection are shown in the top/ bottom panel of Fig. 4b, respectively. The observed PL peak at 22 meV below $D^0$, with a g-factor of -9.8 is consistent with the reported $\Gamma_5$ phonon replica $D^0_{\Gamma_5}$[12,13]. We further note that $D^0_{\Gamma_5}$ shows a zero-field splitting of 0.6 meV, which arises from the fine structure of $D^0$ (Fig. S9)[8,21]. We note that there are two replicas of $D^0$ that are apparent in the magneto-PL (cross-patterns at 3 meV and 13 meV below $D^0$). The peak position of the latter one matches that of positive dark trion thus is precursor of $D^+$, while the former one cannot be unambiguously identified at this time.

The $I^0$ emission is distinct from that of $D^0$ and exhibits all the expected behavior of the momentum indirect, or intervalley dark exciton, which has not been previously identified in this system. This emission peak is strongly co-circularly polarized, with near unity polarization. Its g-factor of about -12.5 corresponds to the recombination of electron and hole residing in opposite valleys, implying that $I^0$ is the direct recombination of intervalley exciton. The formation of $I^0$ is a consequence of the fast valley-flip of electrons via scattering with $K_3$ phonon in the dark exciton formation, as illustrated in Fig. 4c. This is consistent with formation of $D^+(-K)$ discussed above (Fig. 2c). The 10 meV energy splitting between $D^0$ and $I^0$ is then a direct measure of the short-range electron-hole exchange interactions[41]. We note that the direct photon emission from the momentum indirect $I^0$ is weak in intensity, even when compared to the spin-forbidden dark state $D^0$. However, weak PL from indirect states is not unprecedented. Similar to the case of indirect band-gap semiconductors such as silicon, recombination of intervalley excitons without phonons can be mediated by localization next to defects[42], which alleviates the need to conserve crystal momentum due to translation symmetry.

The assignment of $I^0$ is corroborated by the identifications of its $K_1$ and $K_3$ phonon replicas, made evident by examining the energy, polarization, and g-factors of the spectral features. The feature indicated as $I^0_{K_1}$ in Fig. 4a, is located 18 meV below $I^0$, is cross-circularly polarized, and has a positive g-factor of 12.0. These values are nearly identical to those found for both $D^+_{K_1}$ and $D^-_{K_1}$, and consistent with expectations for the $K_1$ valley phonon replica of $I^0$. The $I^0_{K_1}$ recombination process via intervalley hole scattering is illustrated in Fig. 4d. In addition, the peak $I^0_{K_3}$ is about 26 meV below $I^0$ and has a g-factor of -12.6, which is the same as $I^0$ and supports its origin as $K_3$ valley phonon replica of $I^0$ (Fig. 4e).

In conclusion, we unravel the role of valley phonons in exciton and trion formation and their recombination in semiconducting monolayer WSe$_2$. Our work settles questions of the origin of nearly all of the observed peaks in the complex excitonic spectrum of monolayer WSe$_2$. Another important result is that the relaxation of optically generated electrons from the upper conduction band to the lower conduction band is dominated by $K_3$ phonon-assisted spin-conserving intervalley scattering, rather than the $\Gamma_5$ phonon-assisted spin-flip intravalley scattering. Such a relaxation pathway gives rise to the unexpected initial state for the positively charged dark trion, and efficient formation of intervalley exciton $I^0$. This understanding is important for correct interpretation of excitonic spectral features, and may allow for new schemes to control the electron/exciton spin-valley state via optical pumping, e.g. coherent control of $D^0$ and $I^0$ populations via stimulated Raman adiabatic passage. Our work further motivates detailed studies of the electron-valley phonon coupling strengths, which should provide insights for theoretical models to gain a complete understanding of the complex monolayer WSe$_2$ spectrum.

**Methods:**

*Sample Fabrication*
Monolayers of WSe$_2$ were mechanically exfoliated from bulk crystals and identified by optical contrast, which was later confirmed by their low temperature PL spectrum. Thickness of hBN flakes used for encapsulation was typically 10-20nm, while the thickness of graphite back gate electrodes was typically around 5nm. Heterostructures of hBN/WSe2/hBN/Graphite are made with dry-transfer technique using polycarbonate films[43]. The surface of every flake in the heterostructure was confirmed clean with atomic force microscopy prior to fabrication. Finally, the V/Au contact are patterned with standard electron beam lithography and evaporation.

*Photoluminescence Spectroscopy*
PL measurements were performed with a confocal microscope in reflection geometry, with sample mounted in an exchange gas cooled cryostat (AttoDry 2100). The cryostat is equipped with a superconducting magnet in Faraday geometry (magnetic field B perpendicular to sample plane). All measurements were performed at 1.6K unless otherwise specified. A He-Ne laser (632.8 nm) or a frequency tunable continuous-wave Ti:sapphire laser were used to excite the sample. Polarization resolved PL measurements were performed with a set of broad-band half wave plates, quarter wave plates and linear polarizers. PL signal was collected by a spectrometer with a silicon charge-coupled device.

**Acknowledgments:** This work was mainly supported by the Department of Energy, Basic Energy Sciences, Materials Sciences and Engineering Division (DE-SC0018171). Part of understanding of the valley phonon physics is supported by DoE BES DE-SC0014349. Device fabrication and part of magneto optical spectroscopy work are supported by Army Research Office (ARO) Multidisciplinary University Research Initiative (MURI) program, grant no. W911NF-18-1-0431and NSF MRSEC 1719797. WY and HY were supported by Research Grants Council of Hong Kong (17312916), and Seed Funding for Strategic Interdisciplinary Research Scheme of HKU. DM and JY were supported by the US Department of Energy, Office of Science, Basic Energy Sciences, Materials Sciences and Engineering Division. KW and TT were supported by the Elemental Strategy Initiative conducted by the MEXT, Japan and and the CREST (JPMJCR15F3), JST. XX acknowledges the support from the State of Washington funded Clean Energy Institute and from the Boeing Distinguished Professorship in Physics.

**Author Contributions:** MH and PR fabricated the devices and performed the measurements, assisted by NPW and supervised by XX & WY. HD, WY, DVT, MY, HY, MH, PR, and XX analyzed and interpreted the results. JY and DGM synthesized and characterized the bulk WSe$_2$ crystals. KW and TT provided the bulk hBN crystals. MH, PR, XX, HD, and WY wrote the paper with inputs from all authors. All authors discussed the results.

**Competing Financial Interests:** The authors declare no competing financial interests.

**Data Availability**: The data that support the findings of this study are available from the corresponding authors upon reasonable request.

**References:**
1    Xu, X. D., Yao, W., Xiao, D. & Heinz, T. F. Spin and pseudospins in layered transition metal dichalcogenides. *Nat. Phys.* **10**, 343-350 (2014).
2    Xiao, D., Liu, G.-B., Feng, W., Xu, X. & Yao, W. Coupled Spin and Valley Physics in Monolayers of MoS$_2$ and Other Group-VI Dichalcogenides. *Phys. Rev. Lett.* **108**, 196802 (2012).
3    Wang, G. *et al.* Colloquium: Excitons in atomically thin transition metal dichalcogenides. *Rev. Mod. Phys.* **90**, 021001 (2018).
4    Mak, K. F., Xiao, D. & Shan, J. Light–valley interactions in 2D semiconductors. *Nat. Photon.* **12**, 451-460 (2018).
5    Cao, T. *et al.* Valley-selective circular dichroism of monolayer molybdenum disulphide. *Nat. Commun.* **3**, 887 (2012).
6    Manzeli, S., Ovchinnikov, D., Pasquier, D., Yazyev, O. V. & Kis, A. 2D transition metal dichalcogenides. *Nat. Rev. Mater.* **2**, 17033 (2017).
7    Song, Y. & Dery, H. Transport theory of monolayer transition-metal dichalcogenides through symmetry. *Phys. Rev. Lett.* **111**, 026601 (2013).
8    Dery, H. & Song, Y. Polarization analysis of excitons in monolayer and bilayer transition-metal dichalcogenides. *Phys Rev B* **92**, 125431 (2015).
9    Zhang, L. & Niu, Q. Chiral Phonons at High-Symmetry Points in Monolayer Hexagonal Lattices. *Phys. Rev. Lett.* **115**, 115502 (2015).
10   Zhu, H. *et al.* Observation of chiral phonons. *Science* **359**, 579-582 (2018).
11   Chen, X. *et al.* Entanglement of single-photons and chiral phonons in atomically thin WSe$_2$. *Nat. Phys.* **15**, 221-227 (2019).
12   Liu, E. *et al.* Chiral-phonon replicas of dark excitonic states in monolayer WSe$_2$. *arXiv preprint arXiv:1906.02323* (2019).
13   Li, Z. *et al.* Emerging photoluminescence from the dark-exciton phonon replica in monolayer WSe$_2$. *Nat. Commun.* **10**, 2469 (2019).
14   Xiao, D., Yao, W. & Niu, Q. Valley-contrasting physics in graphene: magnetic moment and topological transport. *Phys. Rev. Lett.* **99**, 236809 (2007).
15   Mak, K. F., McGill, K. L., Park, J. & McEuen, P. L. The valley Hall effect in MoS$_2$ transistors. *Science* **344**, 1489-1492 (2014).
16   Gorbachev, R. V. *et al.* Detecting topological currents in graphene superlattices. *Science* **346**, 448-451 (2014).
17   Zhang, X.-X., You, Y., Zhao, S. Y. F. & Heinz, T. F. Experimental Evidence for Dark Excitons in Monolayer WSe$_2$. *Phys. Rev. Lett.* **115**, 257403 (2015).
18   Zhang, X.-X. *et al.* Magnetic brightening and control of dark excitons in monolayer WSe$_2$. *Nat. Nanotech.* **12**, 883 (2017).
19   Zhou, Y. *et al.* Probing dark excitons in atomically thin semiconductors via near-field coupling to surface plasmon polaritons. *Nat. Nanotech.* **12**, 856 (2017).

20	Wang, G. *et al.* In-Plane Propagation of Light in Transition Metal Dichalcogenide Monolayers: Optical Selection Rules. *Phys. Rev. Lett.* **119**, 047401 (2017).
21	Robert, C. *et al.* Fine structure and lifetime of dark excitons in transition metal dichalcogenide monolayers. *Phys Rev B* **96**, 155423 (2017).
22	Tang, Y., Mak, K. F. & Shan, J. Long valley lifetime of dark excitons in single-layer $WSe_2$. *arXiv preprint arXiv:1903.12586* (2019).
23	You, Y. *et al.* Observation of biexcitons in monolayer $WSe_2$. *Nat. Phys.* **11**, 477 (2015).
24	Stier, A. V. *et al.* Magnetooptics of Exciton Rydberg States in a Monolayer Semiconductor. *Phys. Rev. Lett.* **120**, 057405 (2018).
25	Barbone, M. *et al.* Charge-tuneable biexciton complexes in monolayer $WSe_2$. *Nat. Commun.* **9**, 3721 (2018).
26	Li, Z. *et al.* Revealing the biexciton and trion-exciton complexes in BN encapsulated $WSe_2$. *Nat. Commun.* **9**, 3719 (2018).
27	Chen, S.-Y., Goldstein, T., Taniguchi, T., Watanabe, K. & Yan, J. Coulomb-bound four- and five-particle intervalley states in an atomically-thin semiconductor. *Nat. Commun.* **9**, 3717 (2018).
28	Ye, Z. *et al.* Efficient generation of neutral and charged biexcitons in encapsulated $WSe_2$ monolayers. *Nat. Commun.* **9**, 3718 (2018).
29	Jones, A. M. *et al.* Optical generation of excitonic valley coherence in monolayer WSe2. *Nat. Nanotech.* **8**, 634-638 (2013).
30	He, K. *et al.* Tightly Bound Excitons in Monolayer $WSe_2$. *Phys. Rev. Lett.* **113**, 026803 (2014).
31	Ross, J. S. *et al.* Electrically tunable excitonic light-emitting diodes based on monolayer WSe2 p–n junctions. *Nat. Nanotech.* **9**, 268 (2014).
32	Splendiani, A. *et al.* Emerging Photoluminescence in Monolayer $MoS_2$. *Nano Lett.* **10**, 1271-1275 (2010).
33	Mak, K. F., Lee, C., Hone, J., Shan, J. & Heinz, T. F. Atomically Thin $MoS_2$: A New Direct-Gap Semiconductor. *Phys. Rev. Lett.* **105**, 136805 (2010).
34	Park, K.-D., Jiang, T., Clark, G., Xu, X. & Raschke, M. B. Radiative control of dark excitons at room temperature by nano-optical antenna-tip Purcell effect. *Nat. Nanotech.* **13**, 59-64 (2018).
35	Liu, E. *et al.* Gate Tunable Dark Trions in Monolayer $WSe_2$. *Phys. Rev. Lett.* **123**, 027401 (2019).
36	Aivazian, G. *et al.* Magnetic control of valley pseudospin in monolayer $WSe_2$. *Nat. Phys.* **11**, 148 (2015).
37	Srivastava, A. *et al.* Valley Zeeman effect in elementary optical excitations of monolayer $WSe_2$. *Nat. Phys.* **11**, 141 (2015).
38	Lyons, T. P. *et al.* The valley Zeeman effect in inter- and intra-valley trions in monolayer $WSe_2$. *Nat. Commun.* **10**, 2330 (2019).
39	Li, Y. *et al.* Valley Splitting and Polarization by the Zeeman Effect in Monolayer $MoSe_2$. *Phys. Rev. Lett.* **113**, 266804 (2014).
40	MacNeill, D. *et al.* Breaking of Valley Degeneracy by Magnetic Field in Monolayer $MoSe_2$. *Phys. Rev. Lett.* **114**, 037401 (2015).
41	Yu, H., Liu, G.-B., Gong, P., Xu, X. & Yao, W. Dirac cones and Dirac saddle points of bright excitons in monolayer transition metal dichalcogenides. *Nat. Commun.* **5**, 3876 (2014).
42	Karaiskaj, D. *et al.* Photoluminescence of Isotopically Purified Silicon: How Sharp are Bound Exciton Transitions? *Phys. Rev. Lett.* **86**, 6010-6013 (2001).
43	Wang, L. *et al.* One-dimensional electrical contact to a two-dimensional material. *Science* **342**, 614-617 (2013).

# Figures

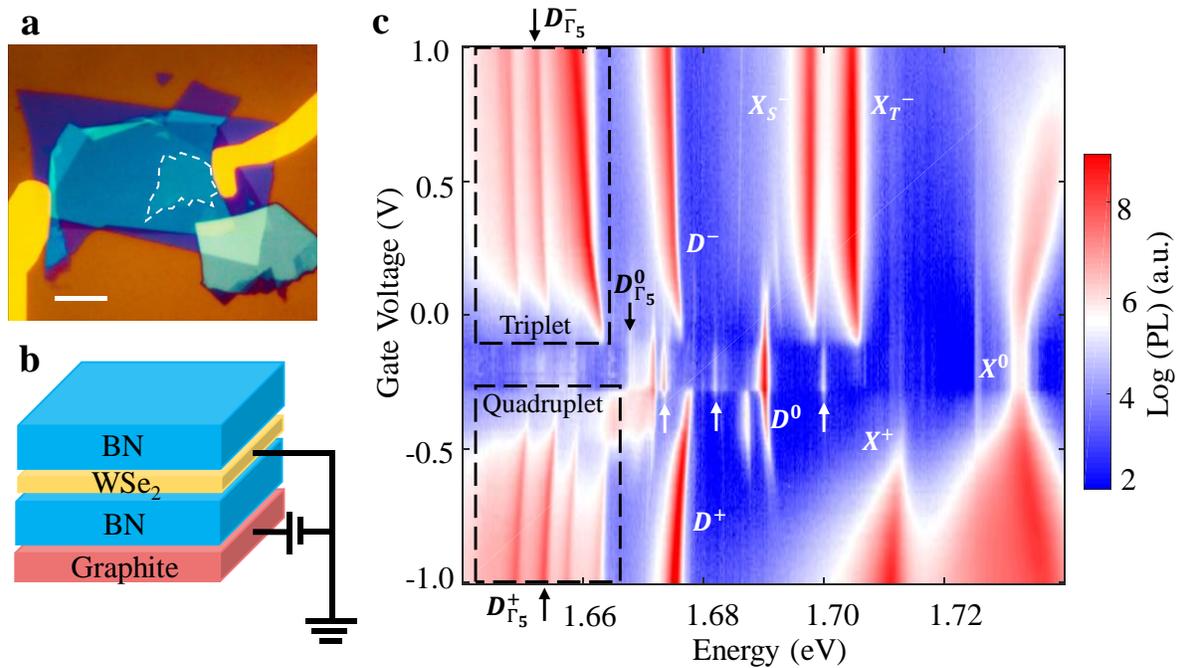

**Figure 1 | Gate dependent photoluminescence. a,** Optical image of a representative gated WSe$_2$ device, in which monolayer WSe$_2$ (white dashed line area) is encapsulated in hBN with a graphite local back gate, scale bar is 10 µm. **b,** Schematic of a gated WSe$_2$ device. **c,** Photoluminescence (PL) as a function of back gate voltage and photon energy. Excitonic states which have been reported in the literature are identified and marked. $X_S^-$ and $X_T^-$ represent the intravalley and intervalley trion. The unidentified triplet PL peaks at electron doping, quadruplet PL peaks at hole doping, and three states at the neutral regime, pointed to by the white arrows, are the focus of this work.

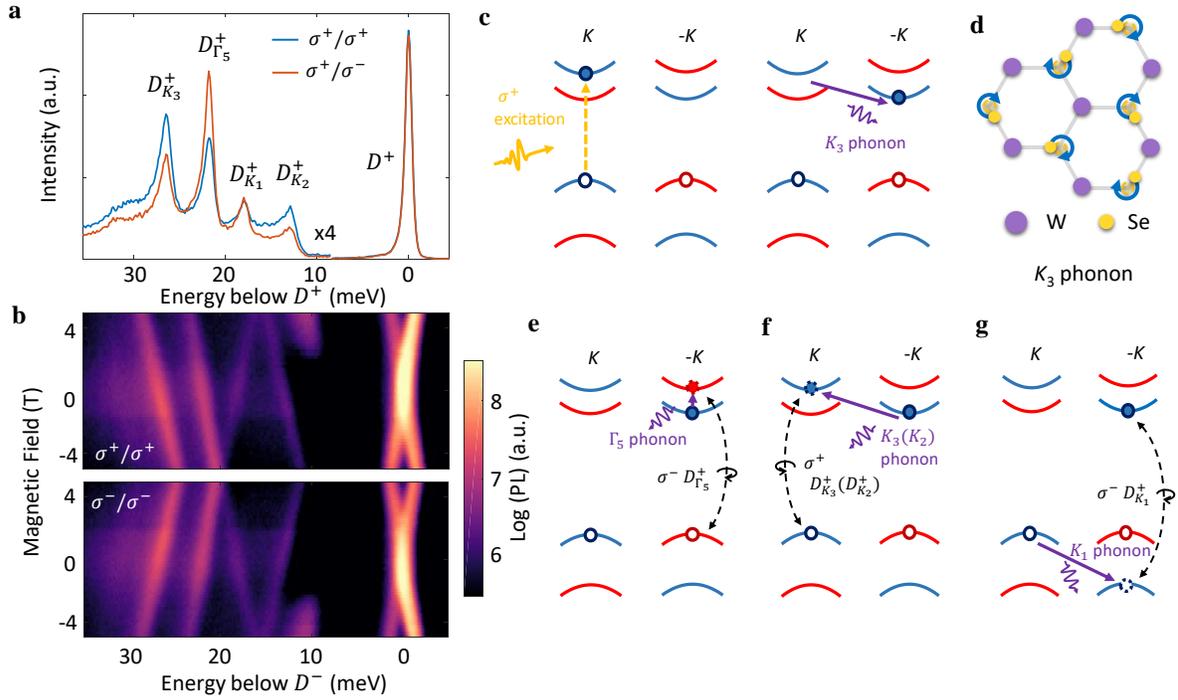

**Figure 2 | Valley phonon replicas of positively charged dark trion. a,** Circular polarization resolved PL of the quadruplet PL peaks on the hole doping side. The photon energy of the spectrum is offset with respect to the positively charged dark trion ($D^+$). The intensity of the quadruplets is multiplied by a factor 4 to emphasize the weak spetrum features. **b,** Circular polarization resolved magneto PL of the quadruplets with $\sigma^+/\sigma^+$ (top) and $\sigma^-/\sigma^-$ (bottom) excitation/detection. **c,** Schematic of $D^+$ formation process under $\sigma^+$ excitation. Blue and red represent bands with electron having spin up and down, respectively. Filled and unfilled circles represent electron and missing electron (hole) in the conduction and valence band, respectively. **d,** Illustration of vibrational normal mode of $K_3$ phonon, wherein Se atoms orbit around their equilibrium positions. See Fig. S6 for other valley phonons. **e-g,** Schematic of phonon-assisted emission process of $D^+_{\Gamma_5}$, $D^+_{K_3}(D^+_{K_2})$, and $D^+_{K_1}$ phonon replicas, respectively. Purple arrow: scattering process with phonons. Dashed circle: virtual state. Black dashed line: light emission from recombination of an electron-hole pair. See text for details.

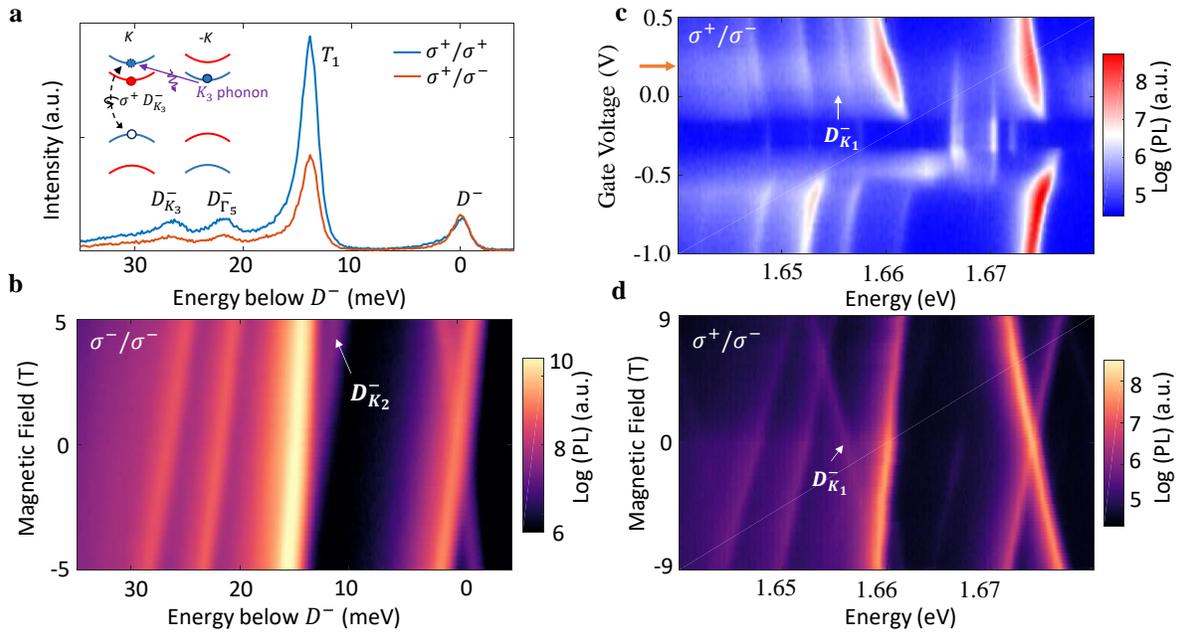

**Figure 3 | Valley phonon replicas of negatively charged dark trion. a,** Circular polarization resolved PL of the triplet PL peaks under electron-doped conditions. The photon energy of the spectrum is offset with respect to the negatively charged dark trion ($D^-$). **b,** Magneto PL of the triplets with $\sigma^-$ excitation and co-polarized detection. The photon energy of the spectrum is offset with respect to $D^-$ at zero field. **c,** Gate dependent PL with $\sigma^+$ resonant pumping of the bright exciton and cross polarized ($\sigma^-$) detection. The $D^-_{K_1}$ state on the electron doping side, which is 18meV below $D^-$, is clearly resolved. **d,** Magneto PL at the gate voltage indicated by the orange arrow in c, with $\sigma^+$ excitation and $\sigma^-$ detection.

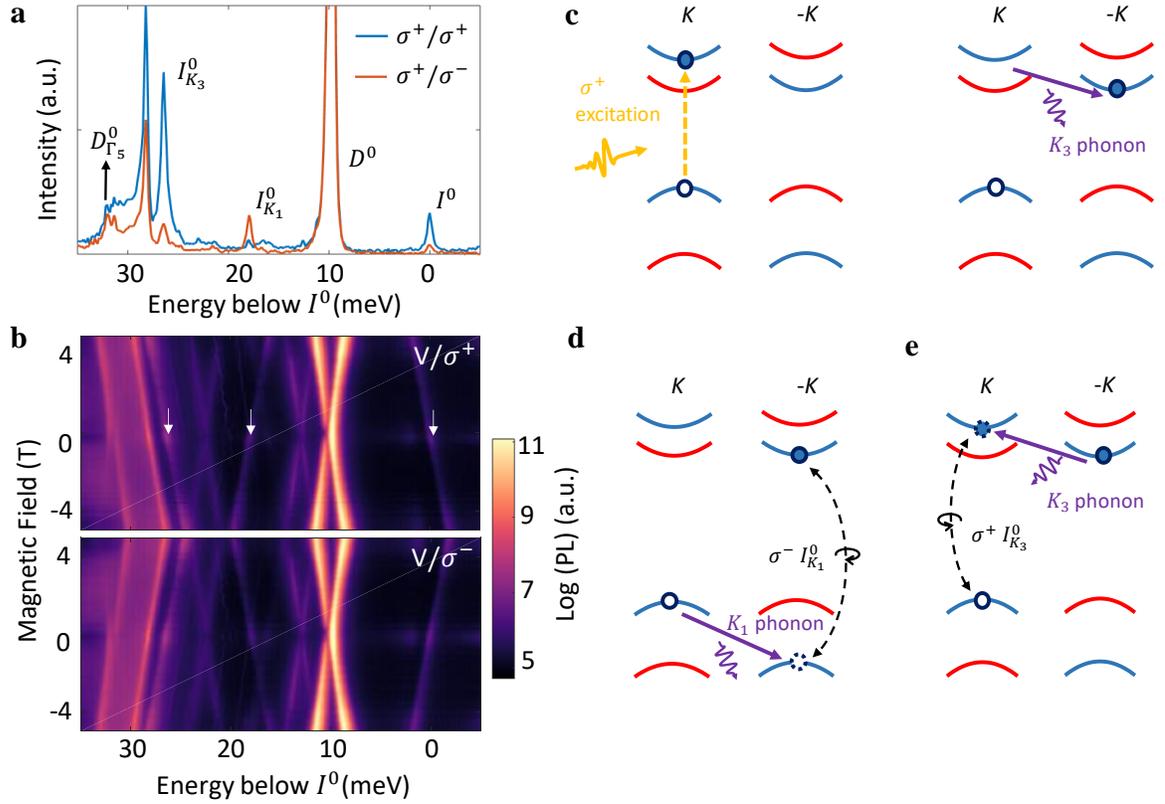

**Figure 4 | Identification of intervalley dark exciton and its valley phonon replicas. a,** Circular polarization resolved PL in the neutral regime. The photon energy of the spectrum is offset with respect to the intervalley exciton ($I^0$). **b,** Magneto PL with linearly polarized excitation, $\sigma^+$ (up) and $\sigma^-$ polarized (down) detection. $I^0_{K_3}$, $I^0_{K_1}$ and $I^0$ are labelled by the white arrows, from left to right. **c,** Schematic of intervalley exciton formation process under $\sigma^+$ excitation, with assistance of $K_3$ phonon. **d,e,** Schematic of light emission process of $I^0_{K_1}$ (left) and $I^0_{K_3}$ (right) phonon replicas.

## Tables

|  |  | Dark state | $K_2$ replica | $K_1$ replica | $\Gamma_5$ replica | $K_3$ replica |
|---|---|---|---|---|---|---|
| neutral regime | $I^0$ | -12.5 | / | 12.0 | / | -12.6 |
|  | $D^0$ | -9.1 | / | / | -9.8 | / |
| Hole doped: $D^+$ |  | -8.6 | -13.0 | 12.2 | -9.7 | -13.4 |
| Electron doped: $D^-$ |  | -9.5 | -13.6 | 12.2 | -9.9 | -12.5 |

**Table 1 | Effective Landé g-factors of WSe$_2$ dark excitonic states and their phonon replicas.**

# Supplementary Materials for
# Valley Phonons and Exciton Complexes in a Monolayer Semiconductor

**Authors:** Minhao He[1†], Pasqual Rivera[1†], Dinh Van Tuan[2], Nathan P. Wilson[1], Min Yang[2], Takashi Taniguchi[3], Kenji Watanabe[3], Jiaqiang Yan[4,5], David G. Mandrus[4-6], Hongyi Yu[7], Hanan Dery[2,8#], Wang Yao[7#], & Xiaodong Xu[1,9#]

**Affiliations:**

[1]Department of Physics, University of Washington, Seattle, Washington 98195, US A

[2]Department of Electrical and Computer Engineering, University of Rochester, Rochester, New York, 14627, USA

[3]National Institute for Materials Science, Tsukuba, Ibaraki 305-0044, Japan

[4]Materials Science and Technology Division, Oak Ridge National Laboratory, Oak Ridge, Tennessee, 37831, USA

[5]Department of Materials Science and Engineering, University of Tennessee, Knoxville, Tennessee, 37996, USA

[6]Department of Physics and Astronomy, University of Tennessee, Knoxville, Tennessee, 37996, USA

[7]Department of Physics and Center of Theoretical and Computational Physics, University of Hong Kong, Hong Kong, China

[8]Department of Physics and Astronomy, University of Rochester, Rochester, New York 14627, USA

[9]Department of Materials Science and Engineering, University of Washington, Seattle, Washington, 98195, USA

[†] These authors contributed equally to the work.

[#]Correspondence to xuxd@uw.edu; wangyao@hku.hk; hanan.dery@rochester.edu

**Table of Contents**

**List of figures and tables**

## SI-1. Reproducible gate dependent photoluminescence

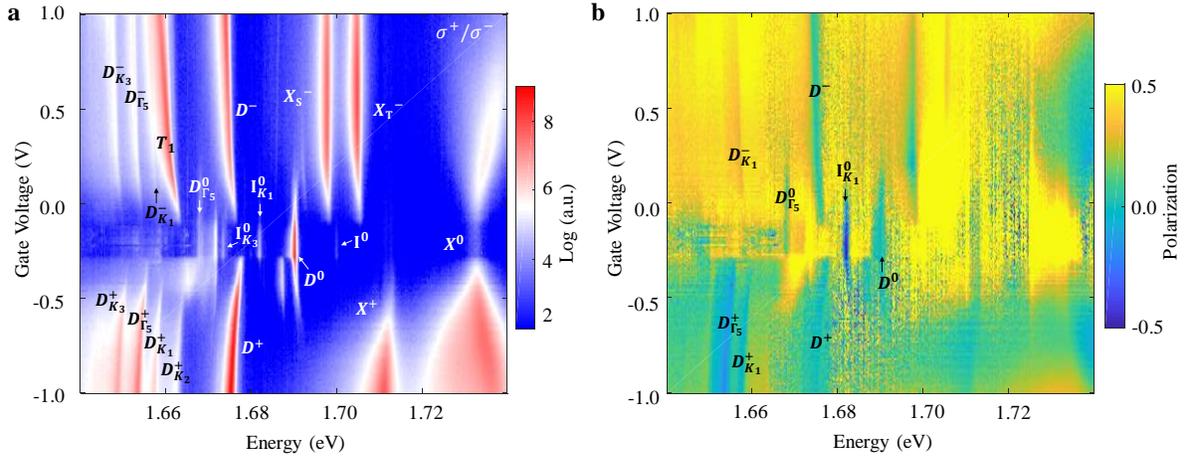

**Figure S1| Gate dependent photoluminescence. a,** PL intensity plot as a function of gate voltage and photon energy, from the same device as shown in Fig. 1c. The excitation and detection are cross circularly polarized ($\sigma^+$ excitation and $\sigma^-$ detection). We marked the excitonic states that have been identified in this paper as well as previously reported. **b,** The degree of circular of polarization as a function of gate voltage. $D^-$, $D^0$ and $D^+$ have negligible circular polarization due to their out of plane dipole moment. $I^0_{K_1}$ and $D^+_{\Gamma_5}$ stand out for their obvious cross polarization.

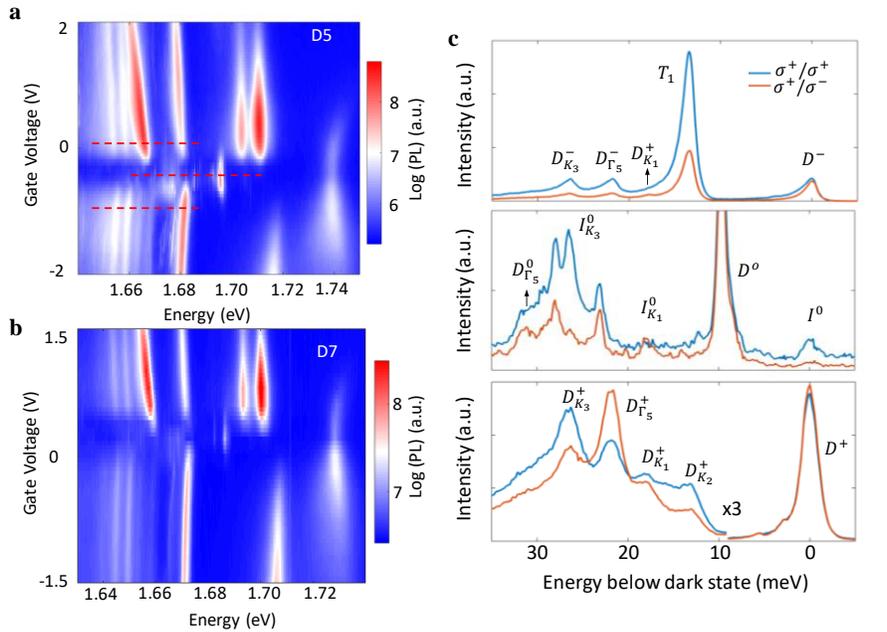

**Figure S2| Reproducible gate dependent spectrum. a,b,** Gate dependent PL spectrum in additional devices. Both devices show similar spectrum as in Fig. 1c, with well resolved valley phonon replicas. **c,** Circular polarization resolved PL in MD 5, at selective gate voltages indicated by the red lines with electron doping, neutral regime and hole doping from top to bottom. Valley phonon replicas and the dark excitonic states are labeled for clarity. The photon energy of the spectrum is relative to the corresponding indirect exciton ($I^0$) or dark trions ($D^\pm$).

## SI-2. Power dependence of valley phonon-assisted emission.

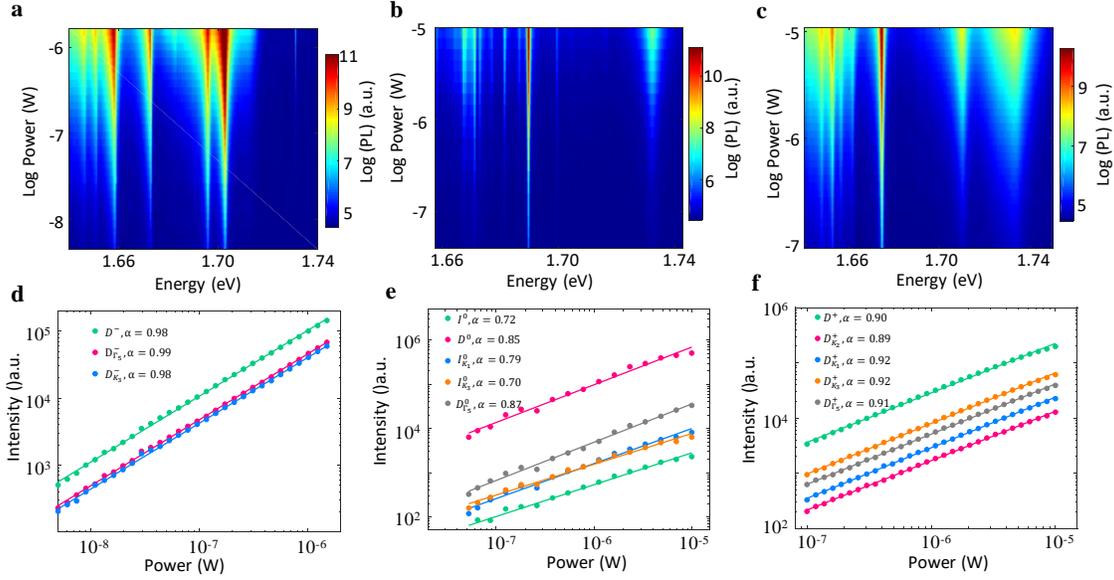

**Figure S3| Power dependence of valley phonon-assisted emission. a-c,** PL intensity as a function of pumping power, at electron doping, neutral regime and hole doping from left to right. **d-f,** Extracted intensity of dark states and their phonon replicas as a function of pumping power. The solid lines are power law fitting. For each dark state, the corresponding power dependence of valley phonon replicas are all similar. This further supports their valley assisted emission from the same dark state.

## SI-3. Landé effective g-factor analysis.

In monolayer WSe$_2$, Zeeman splitting of excitonic states at the ±$K$-point valleys can be attributed to three main contributions[1-4]. The first part comes from spin of the composite quasiparticles in the excitonic state, which gives a Zeeman energy of $2S_z\mu_B B$. The second part comes from the atomic orbital magnetic moment, giving a Zeeman energy of $m\mu_B B$. In the case of monolayer WSe2, the conduction band edge is mainly composed of d-orbitals with m = 0 in both K-point valleys, whereas the valence band edges in the $K$(-$K$) point valley are mainly d-orbitals with $m$ = 2(-2) in the upper valence band, and $m = -2(2)$ in the lower valence band. Lastly, it is shown that the lattice structure also contributes a valley magnetic moment with a g-factor of $\alpha_c(\alpha_v)$ for the conduction(valence) band, resulting in a Zeeman energy of $\alpha_c(\alpha_v) \mu_B B$ for an electron (hole), respectively.

Fig. S4a shows three main contributions to the Zeeman shift at the band edges in ±$K$ point valleys. We identify three representative spin-valley configurations involved in the light emission process of all the excitonic states in Fig. S1a. Notice that for the trion states, the quasiparticle that remain unchanged after light emission does not affect the measured Zeeman splitting. Thus, we only consider spin-valley configuration of the electron-hole pair involved in the recombination. Intravalley bright exciton recombination, shown in Fig. S4b, refers to recombination of an electron in the conduction band and a missing electron (hole) in the valence band with the same spin and valley quantum numbers. The calculated effective g-factor is $2(\alpha_c-\alpha_v-2)$, and is experimentally measured to be about -4. This means the valley magnetic moments for the conduction and valence bands are almost the same, which agrees well with theory[3-6].

Intravalley dark recombination refers to recombination of an electron and a missing electron (hole) in the same valley with opposite spin quantum number, as shown in Fig. S4c. Similarly, we can calculate that the effective g-factor is $2(\alpha_c-\alpha_v-4)$. The corresponding $D^0$, $D^+$ and $D^-$ states have an effective g-factor close to -9[7-9]. Finally, the intervalley recombination refers to recombination of an electron and a missing electron (hole) in the opposite valleys, but with the same spin quantum number. The effective g-factor can be written as $2(-\alpha_c-\alpha_v-4)$ for $K_3$ and $K_2$ valley phonon assisted intervalley recombination. It has the same amplitude but opposite sign for $K_1$ valley phonon assisted intervalley recombination. The complication of the sign of g-factors comes from the circular polarization of the emitted photon, as a result of it being either an intervalley electron scattering process or intervalley hole scattering process. The magnitude of the effective g-factor for the intervalley recombination is measured be about 13.

In Table S1, we summarized the measured effective g-factor of the excitonic states identified in Fig. S1a. We extract the $\sigma^+$ and $\sigma^-$ polarized PL peak energy, $E(\sigma^+)$ and $E(\sigma^-)$, respectively. The effective g-factor is then calculated from the extracted Zeeman splitting: $\Delta = E(\sigma^+) - E(\sigma^-)$.

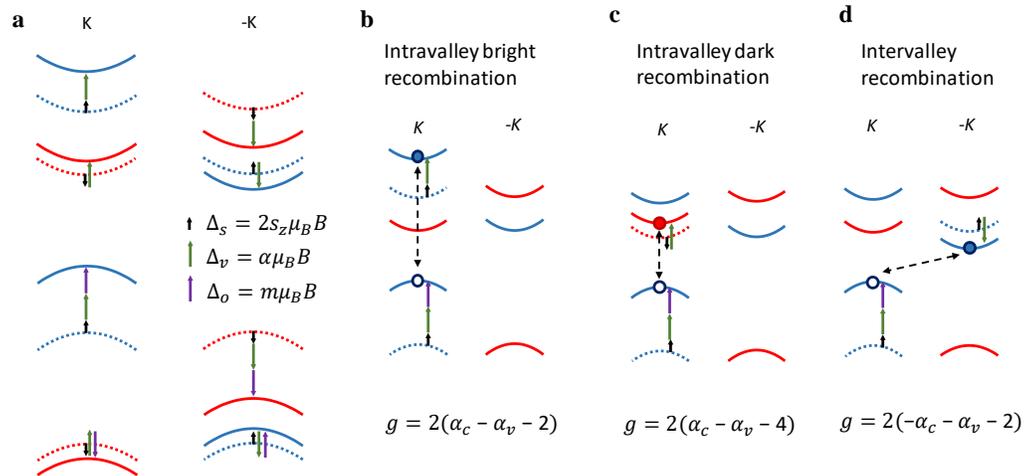

**Figure S4 | Effective Landé g-factor of excitonic states with different spin-valley configurations. a,** Band diagram at $\pm K$ valley, showing three main contributions to Zeeman shifts: black for spin, green for valley magnetic moment, and purple for atomic orbital magnetic moment. **b-d,** Different spin-valley configuration of the recombined electron-hole pair, showing intravalley bright exciton recombination, intravalley dark exciton recombination, and intervalley recombination. Their corresponding effective g-factors are indicated at the bottom.

|  | Bright state | Dark state |  | $K_2$ replica | $K_1$ replica | $\Gamma_5$ replica | $K_3$ replica |
|---|---|---|---|---|---|---|---|
| neutral regime | -4.7 | $I^0$ | -12.5 | / | 12.0 | / | -12.6 |
|  |  | $D^0$ | -9.1 | / | / | -9.8 | / |
| Hole doped: $D^+$ | -2 | | -8.6 | -13.0 | 12.2 | -9.7 | -13.4 |
| Electron doped: $D^-$ | $X_T^-$ -5.2 $X_S^+$ -3.4 | | -9.5 | -13.6 | 12.2 | -9.9 | -12.5 |

**Table S1| Landé effective g-factor of identified excitonic states.** Effective g-factor of identified excitonic states in Fig. S1a. For readers' convenience, g-factor of dark states and their phonon replicas in Table 1 are listed here as well.

**SI-4. Selection rules for intervalley transitions.**

**Phonon-induced intravalley transitions between bright and dark excitons**

Table S2 shows the character table of $D_{3h}$ point double-group[10]. This table is used to derive selection rules for phonon-mediated transitions of intravalley excitons, corresponding to the case that both electron and hole reside in the same valley (e.g., both in the $K$-point valley)[11]. The transformation properties of optically-inactive dark excitons are captured by the irreducible representation (IR) $\Gamma_3$, semi-dark excitons with out-of-plane optical transition dipole by $\Gamma_4$, and bright excitons with in-plane optical transition dipole by $\Gamma_6$[12]. From Table S2, one can readily verify the selection rule,

$$(\Gamma_6^* \times \Gamma_3)^* = (\Gamma_6^* \times \Gamma_4)^* = \Gamma_5, \qquad (S1)$$

implying that transitions between dark and bright excitons due to intravalley spin flip (of the electron component) can be induced by a zone-center phonon that transforms like the IR $\Gamma_5$. This phonon is the homopolar optical mode, whose polarization vector is denoted by in-plane and out-of-phase vibration of the chalcogen atoms, as shown in Fig. S5a[12]. Note that the spin-flip matrix element due to interaction with a long-wavelength flexural phonon, which transforms as $\Gamma_4$, is nonzero only if the phonon wavevector is finite (i.e., when $q \neq 0$)[11]. On the other hand, the phonon mode $\Gamma_5$ is the only one for which the spin-flip matrix element is nonzero exactly at the zone center ($q = 0$).

**Zone-edge phonons**

Using Quantum Expresso[13], Fig. S5b shows the calculated phonon spectrum in monolayer WSe$_2$. Because the concept of acoustic and optical phonon modes near the $\Gamma$-point loses its meaning when $qa$ is no longer much smaller than unity ($a$ is the lattice constant), the notion of acoustic or optical phonons for longitudinal and transverse modes (i.e., LA, TA, LO and TO) is not valid when dealing with zone-edge phonons at the $K$-point whose wavenumber is $q = K = 4\pi/3a$. Instead, the use of $K$-point IRs is more appropriate. This nomenclature is presented in Table S3, where the transformation properties of the zone-edge phonons are captured by the IRs $K_1$ through $K_6$ (these IRs also belong to the $C_{3h}$ point single group. The spin quantum number has no bearing on the polarization vectors of atoms in the unit cell)[11].

| $D_{3h}$ | | $E$ | $\bar{E}$ | $C_3^+ C_3^-$ | $\bar{C}_3^+ \bar{C}_3^-$ | $\sigma_h \bar{\sigma}_h$ | $S_3^+ S_3^-$ | $\bar{S}_3^+ \bar{S}_3^-$ | $C'_{2i} \bar{C}'_{2i}$ | $\sigma_{vi} \bar{\sigma}_{vi}$ |
|---|---|---|---|---|---|---|---|---|---|---|
| $A'_1$ | $\Gamma_1$ | 1 | 1 | 1 | 1 | 1 | 1 | 1 | 1 | 1 |
| $A'_2$ | $\Gamma_2$ | 1 | 1 | 1 | 1 | 1 | 1 | 1 | -1 | -1 |
| $A''_1$ | $\Gamma_3$ | 1 | 1 | 1 | 1 | -1 | -1 | -1 | 1 | -1 |
| $A''_2$ | $\Gamma_4$ | 1 | 1 | 1 | 1 | -1 | -1 | -1 | -1 | 1 |
| $E''$ | $\Gamma_5$ | 2 | 2 | -1 | -1 | -2 | 1 | 1 | 0 | 0 |
| $E'$ | $\Gamma_6$ | 2 | 2 | -1 | -1 | 2 | -1 | -1 | 0 | 0 |
| $\bar{E}_1$ | $\Gamma_7$ | 2 | -2 | 1 | -1 | 0 | $\sqrt{3}$ | $-\sqrt{3}$ | 0 | 0 |
| $\bar{E}_2$ | $\Gamma_8$ | 2 | -2 | 1 | -1 | 0 | $-\sqrt{3}$ | $\sqrt{3}$ | 0 | 0 |
| $\bar{E}_3$ | $\Gamma_9$ | 2 | -2 | 2 | 2 | 0 | 0 | 0 | 0 | 0 |

**Table S2. Character table of $D_{3h}$ point double-group.** It captures the symmetry properties of the $\Gamma$ point in ML-TMDs. The $x$-axis is along zigzag edge direction and the $y$-axis is along armchair direction.

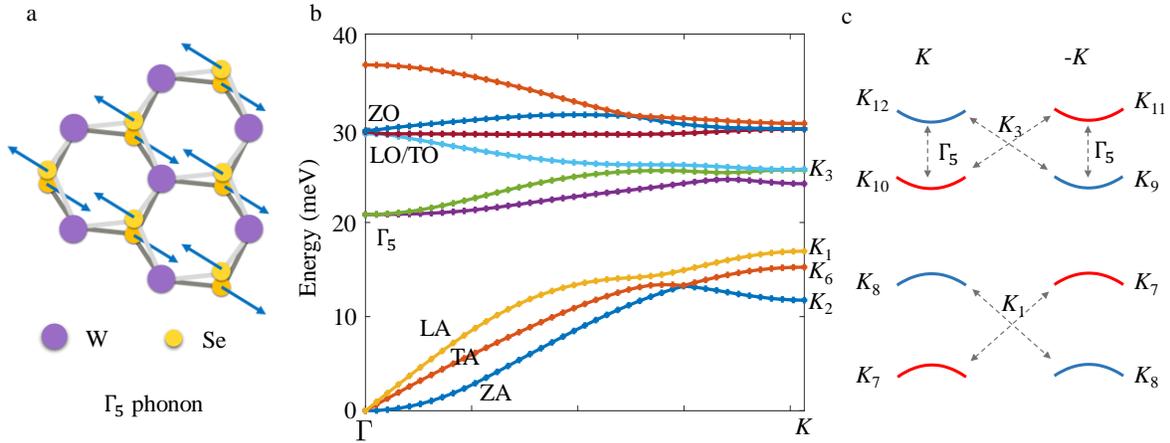

**Figure S5 | Phonon modes and selection rules. a,** Atomic displacement of the phonon mode $\Gamma_5$. **b,** The phonon dispersion of monolayer WSe$_2$ along the axis between the high-symmetry $\Gamma$ and $K$ points. Zone-center and pertinent zone-edge phonon modes are indicated (symmetries and atomic displacements of the zone-edge phonon modes are analyzed in Table S3 and Fig. S6). **c,** Scheme of low-energy valleys in the conduction and valence bands. Spin-flip intravalley transitions in the conduction band are mediated by the phonon-mode $\Gamma_5$, while spin-conserving intervalley transitions in the conduction (valence) bands are mediated by the phonon mode $K_3$ ($K_1$). Phonon-induced intervalley spin-flip scattering is relatively weak (the transition matrix element between time-reversed states vanishes).

The nine polarization vectors of the zone-edge phonons are presented in the last column of Table S3. $M_{z/\parallel}(x,y,z)$ and $X_{\pm z/\parallel}(x,y,z)$ are polarization vectors for the transition-metal and chalcogen atoms, respectively. The subscripts $z$ and $\parallel$ denote out-of plane ($z$) and in-plane ($\parallel$) vibrations. The subscript '+(−)' of X denotes in(counter)-phase motion of the two chalcogen atoms in a unit cell. Table S3 shows that each of the IRs $K_{3-5}$ is associated with one polarization vector, and consequently, each of these IRs represents one zone-edge phonon. The IRs $K_1$, $K_2$ and $K_6$ are different in that each contains two types of atomic displacements, and thus, each corresponds to two independent zone-edge phonon modes. From energy considerations that will be explained below and using Fig. S5b as a typical example for ML-TMDs, $K_{3-5}$ represent zone-edge phonons from the mid-three bundle whereas the bottom (top) three zone-edge phonons are associated with the low- (high-) energy modes of $K_1$, $K_2$ and $K_6$. The atomic displacement that corresponds to the mode $K_l$ with energy $\hbar\omega_{Kl}$ is proportional to

$$\delta R^{K_1}_{\alpha,j} \propto V_\alpha(K_l)e^{i(K\cdot R_{\alpha,j} + \omega_{K_1}t)} + V_\alpha^*(K_l)e^{-i(K\cdot R_{\alpha,j} + \omega_{K_1}t)}. \tag{S2}$$

| $C_{3h}$ | | E | $C_3^+$ | $C_3^-$ | $\sigma_h$ | $S_3^+$ | $S_3^-$ | $\bar{E}$ | $\overline{C_3^+}$ | $\overline{C_3^-}$ | $\overline{\sigma_h}$ | $\overline{S_3^+}$ | $\overline{S_3^-}$ | Polarization vectors |
|---|---|---|---|---|---|---|---|---|---|---|---|---|---|---|
| $A'$ | $K_1$ | 1 | 1 | 1 | 1 | 1 | 1 | 1 | 1 | 1 | 1 | 1 | 1 | $M_\parallel(1,-i,0)$ $\pm X_{+,\parallel}(1,i,0)$ |
| $^2E'$ | $K_2$ | 1 | $\omega$ | $\omega^*$ | 1 | $\omega$ | $\omega^*$ | 1 | $\omega$ | $\omega^*$ | 1 | $\omega$ | $\omega^*$ | $X_{-,z} \pm M_- \parallel (1,i,0)$ |
| $^1E'$ | $K_3$ | 1 | $\omega^*$ | $\omega$ | 1 | $\omega^*$ | $\omega$ | 1 | $\omega^*$ | $\omega$ | 1 | $\omega^*$ | $\omega$ | $X_{+,\parallel}(1,-i,0)$ |
| $A''$ | $K_4$ | 1 | 1 | 1 | -1 | -1 | -1 | 1 | 1 | 1 | -1 | -1 | -1 | $X_{-,\parallel}(1,i,0)$ |
| $^2E''$ | $K_5$ | 1 | $\omega$ | $\omega^*$ | -1 | $-\omega$ | $-\omega^*$ | 1 | $\omega$ | $\omega^*$ | -1 | $-\omega$ | $-\omega^*$ | $X_{+,z}$ |
| $^1E''$ | $K_6$ | 1 | $\omega^*$ | $\omega$ | -1 | $-\omega^*$ | $-\omega$ | 1 | $\omega^*$ | $\omega$ | -1 | $-\omega^*$ | $-\omega$ | $M_z \pm X_{-,\parallel}(1,-i,0)$ |
| $^1\bar{E}_3$ | $K_7$ | 1 | $-\omega$ | $-\omega^*$ | $i$ | $-i\omega$ | $i\omega^*$ | -1 | $\omega$ | $\omega^*$ | $-i$ | $i\omega$ | $-i\omega^*$ | |
| $^2\bar{E}_3$ | $K_8$ | 1 | $-\omega^*$ | $-\omega$ | $-i$ | $i\omega^*$ | $-i\omega$ | -1 | $\omega^*$ | $\omega$ | $i$ | $-i\omega^*$ | $i\omega$ | |
| $^2\bar{E}_2$ | $K_9$ | 1 | $-\omega$ | $-\omega^*$ | $-i$ | $i\omega$ | $-i\omega^*$ | -1 | $\omega$ | $\omega^*$ | $i$ | $-i\omega$ | $i\omega^*$ | |
| $^1\bar{E}_2$ | $K_{10}$ | 1 | $-\omega^*$ | $-\omega$ | $i$ | $-i\omega^*$ | $i\omega$ | -1 | $\omega^*$ | $\omega$ | $-i$ | $i\omega^*$ | $-i\omega$ | |
| $^1\bar{E}_1$ | $K_{11}$ | 1 | -1 | -1 | $i$ | $-i$ | $i$ | -1 | 1 | 1 | $-i$ | $i$ | $-i$ | |
| $^2\bar{E}_1$ | $K_{12}$ | 1 | -1 | -1 | $-i$ | $i$ | $-i$ | -1 | 1 | 1 | $i$ | $-i$ | $i$ | |

**Table S3 | Character table of $C_{3h}$ point double-group.** This table captures the symmetry properties of the $K$ point in ML-TMDs where $\omega = \exp(2\pi i/3)$. To prevent confusion with the notation in Table S2, the Koster notation of the IRs is changed from $\Gamma_i$ to $K_i$. The x-axis is along the zigzag edge direction and the y-axis is along the armchair direction.

$R_{\alpha,j}$ is a 2D vector denoting the equilibrium position of the $\alpha$-atom in the $j^{th}$ unit cell, where $\alpha = (M, X_t, X_b)$ and $X_{t/b}$ denotes the top/bottom chalcogen atoms. $V_\alpha(K_l)$ is the polarization vector (right column of Table S3). Given that $|K| = 4\pi/3a$ where $a$ is the in-plane distance between nearby identical atoms, the phase difference obeys $K \cdot (R_{X(M),j} - R_{X(M),j+n}) = \pm 2\pi n/3$, where $n$ is the number of unit cells along a zig-zag chain. Consequently, the displacement of every third chalcogen (or transition-metal) atom along a zigzag chain is identical. In addition, when substituting the in-plane polarization vectors $V_\alpha = [1,\pm i,0]$ in Eq. (S2), the resulting atomic motion follows a clockwise or counterclockwise circular trajectory, $\delta R^{K_1}_{\alpha,j} \propto \cos(K \cdot R_{\alpha,j} + \omega_{K_1}t)\hat{x} \pm \sin(K \cdot R_{\alpha,j} + \omega_{K_1}t)\hat{y}$.

Supported by our experimental findings and the group theory analysis below, we will focus in this work on the modes $K_{1-3}$. Figure S6a shows the atomic displacements that correspond to low- and high- energy phonon modes with symmetry $K_1$. The difference between the two is the phase difference in the circular motion of Se and W atoms, leading to mitigated deviation of the bond lengths from equilibrium in the low energy case. Figure S6b shows the atomic displacement that corresponds to a phonon mode with symmetry $K_3$. Here, only Se atoms go through in-plane circular motion around their equilibrium positions and the phonon energy is somewhere between that of the low- and high-energy phonon modes with symmetry $K_1$.

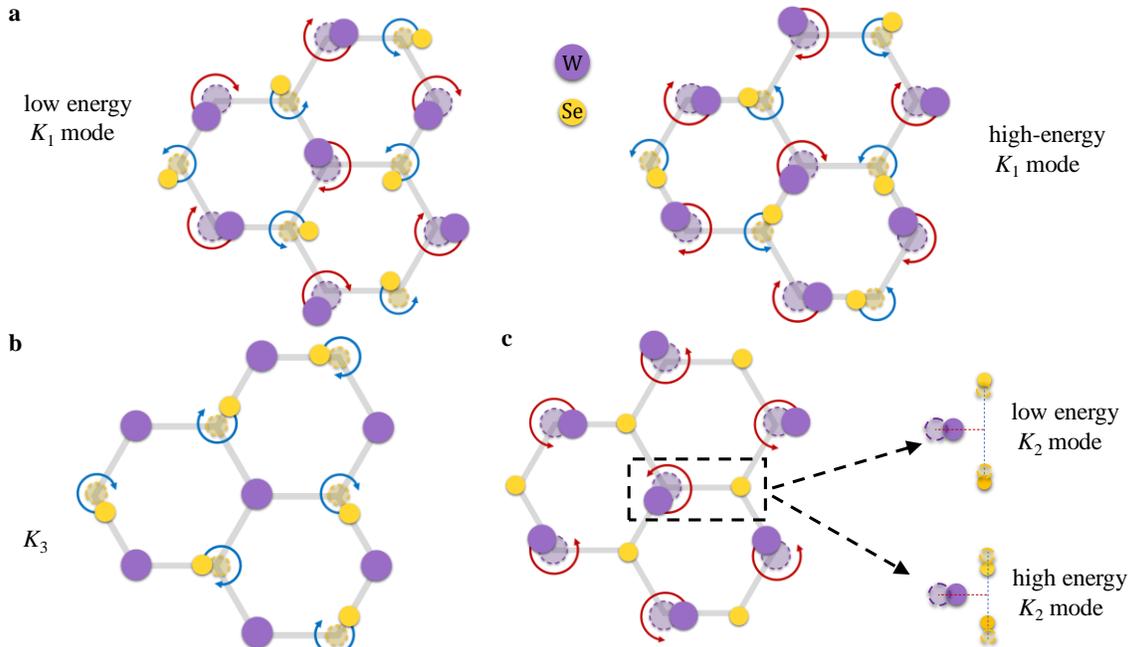

**Figure S6| Atomic motions of Valley phonons.** The curved arrows denote the in-plane circular motion of atoms around their equilibrium positions, denoted by faint circles. See text for further explanations.

Figure S6c shows the atomic displacements that correspond to low- and high- energy phonon modes with symmetry $K_2$. Here, the transition-metal atoms go through in-plane circular motion around their equilibrium positions, while chalcogen atoms vibrate in opposite directions along the out-of-plane axis. As shown in the figure, the difference between the low and high energy modes is the relative motion of the chalcogen and transition-metal atoms. Namely, the low energy mode corresponds to the case that when the transition-metal atoms move toward the chalcogen atoms, the chalcogen atoms move away from the mid-plane and vice versa. This combined motion keeps the bond length closer to its equilibrium value, and hence to lower phonon energy. Conversely, the high energy mode leads to stronger deviations from equilibrium because when the transition-metal atoms move toward the chalcogen atoms, the chalcogen atoms move closer to the mid-plane and vice versa.

Finally, Fig. S5b shows that the branch extensions of the ZA and TA modes anti-cross close to the $K$-point. This anti-crossing is not seen in the calculation of monolayer $MoS_2$. The result is that the mode $K_6$ has the lowest energy in monolayer $MoS_2$, while $K_2$ has the lowest energy in monolayer $WeS_2$.

**Band-edge electronic states and selection rules for intervalley transitions**

Next, we derive selection rules for intervalley transitions of electrons and holes. We first note that $K = 4\pi/3a$ is not only the wavenumber that connects each of the zone-edge $\pm K$-points with the zone center, but it is also the wavenumber needed to connect the zone-edge $+K$ and $-K$ points. In other words, the conservation of crystal momentum due to intervalley transitions of electrons (or holes) states at the $\pm K$ points is mediated through the zone-edge $K$-point phonons. We use Table S3 to find the transformation properties of the zone-edge electronic states along with the ensuing selection rules for phonon-induced intervalley transitions of electrons and holes[11]. Starting with holes states, the transformation properties of their wavefunctions at the valence-band edge states are captured by the IRs $K_7$ and $K_8$. As shown in Fig. S5c, the band-edge hole state at the top $-K$ ($+K$) valley transforms like the IR $K_7$ ($K_8$), and vice versa for the bottom valleys. This behavior is a consequence of time-reversal symmetry and the complex conjugation of the IRs $K_7$ and $K_8$. Accordingly, a spin conserving intervalley scattering between the edge states involves the selection rule[11]

$$(K_7^* \times K_7)^* = (K_8^* \times K_8)^* = K_1, \tag{S3}$$

where the first term corresponds to hole transition between the top valley in $-K$ and the bottom one in $+K$, while the second term corresponds to hole transition between the bottom valley in $-K$ and the top one in $+K$. This selection rule implies that the phonon mode $K_1$ is the dominant mechanism for intervalley transitions in the valence band.

The case of conduction-band electrons is somewhat more subtle. As shown in Fig. S5c, the band-edge electron state at the bottom (top) $-K$ valley transforms like $K_9$ ($K_{11}$), while the respective complex conjugate IRs at the $+K$ valleys transforms like $K_{10}$ ($K_{12}$). The selection rule in this case reads[11]

$$(K_{12}^* \times K_9)^* = (K_{10}^* \times K_{11})^* = K_3, \tag{S4}$$

implying that the phonon mode $K_3$ is the dominant mechanism for intervalley transitions in the conduction band.

**SI-5. Qualitative comparison between theory and experiment**

Analyzing the experimental findings of this work, we find very good agreement with the above theoretical analysis and previous predictions[11,12]. For example, Fig. 4 shows two phonon replicas of the indirect (intervalley) exciton, $I^0$, whose energies match the calculated values of the phonons modes $K_1$ (17 meV) and $K_3$ (26 meV) in Fig. S5b. These modes are suggested by the selection rules in Eqs. (S3) and (S4), and they are supported by the cross-polarized emitted light from the phonon replica $I^0_{K_1}$ versus the co-polarized emitted light from the phonon replica $I^0_{K_3}$, as shown and analyzed in the main text. Similarly, we see that this behavior is consistent when dealing with dark trions states (Fig. 2 and Fig. 3).

One aspect of the experiment that can be further supported by theory is the observation that the PL phonon-assisted peaks associated with the mode $K_3$ are noticeably stronger in amplitude than those with $K_1$. Invoking second-order perturbation theory[14], the amplitude ratio of these peaks follows

$$\frac{A_{K_3}}{A_{K_1}} = \left|\frac{\mathcal{M}_{\mathcal{K}_3}}{\mathcal{M}_{\mathcal{K}_1}}\right|^2 \times \left|\frac{E_{D^\pm} - \left(E_{X_B^\pm} + E_{K_1}\right)}{E_{D^\pm} - \left(E_{X_A^\pm} + E_{K_3}\right)}\right|^2. \tag{S5}$$

$\mathcal{M}_{\mathcal{K}_3}$ and $\mathcal{M}_{\mathcal{K}_1}$ are the matrix elements for phonon-mediated intervalley transition in the conduction and valence bands, respectively. The other terms in Eq. (S5) are the energy differences between the initial state, here assumed to be the positive or negative dark trion ($D^{\pm}$), and the intermediate virtual states. For the case of intervalley hole transition, the energy of the intermediate state is that of the emitted phonon $K_1$ and the virtual type-B trion. For the case of intervalley electron transition, the energy of the intermediate state is that of the emitted phonon $K_3$ and the virtual bright type-A trion state (with singlet spin configuration of the same-sign charges).

The reason for the stronger signature of $K_3$ phonons compared with $K_1$ is the small spin-splitting energy of the conduction band compared with that of the valence-band, $\Delta_c \ll \Delta_v$. As a result, the energies of the initial and intermediate virtual states are relatively similar when the intermediate hole states are kept in the top valleys of the valence bands (i.e., when the electron goes through intervalley transition whereas the hole is a spectator). Substituting empirical values for the energies in monolayer WSe$_2$, $|E_{D^{\pm}} - (E_{X_A^{\pm}} + E_{K3})| \sim 60$ meV, and $|E_{D^{\pm}} - (E_{X_B^{\pm}} + E_{K1})| \sim \Delta_v \sim 400$ meV, we get that

$$\frac{A_{K_3}}{A_{K_1}} \sim 40 \left|\frac{\mathcal{M}_{\mathcal{K}_3}}{\mathcal{M}_{\mathcal{K}_1}}\right|^2. \tag{S6}$$

Our experimental results show that the amplitude ratio of the peaks associated with $K_3$ and $K_1$ is about 4, implying that $\mathcal{M}_{\mathcal{K}_1}$ should be about three times larger than $\mathcal{M}_{\mathcal{K}_3}$. This empirical analysis can be used to benchmark the results of future first-principles calculations of these matrix elements.

**SI-6. The phonon replica K$_2$**

In addition to the replicas predicted by the previous group-theory analysis, we have observed a weak phonon replica in both the positive and negative dark trions that emerges ~13 meV below the no-phonon dark trion lines (Fig.2 and Fig. 3). From the *g*-factor and polarization analysis, one can see that this replica, albeit weaker, shows the same characteristics of the replica $K_3$. Namely, it involves intervalley scattering of the electron component. A question remains regarding the physical origin of this replica. Inspecting the low-energy modes at the *K*-point, as shown in Fig. S5b, we can associate these peaks with either the lowest-energy mode $K_2$ or the second one $K_6$ (their calculated energies are 11.7 and 15.3 meV).

The mode $K_6$ involves out-of-plane vibration of the transition-metal atom, which breaks the mirror-inversion symmetry of the monolayer. In general, interactions of electrons (or holes) with such phonon modes can only couple to spin flips[11]. Unlike the zone-edge phonons with $K_6$ symmetry in which only W atoms vibrate in the out-of-plane direction, zone-edge phonons with $K_2$ symmetry involve counter motion of the Se atoms in the out-of-plane direction which retains the mirror inversion symmetry. Accordingly, the mode $K_2$ can induce a spin-conserving intervalley transition, albeit its amplitude is measurable only when the initial and final states are relatively far from being time reversal partners. In more detail, the selection rule in Eq. (S4) does not only mean that the amplitude of the matrix element $\mathcal{M}_{\mathcal{K}_3}$ is the dominant one for spin-conserving intervalley electron transitions, but that it is the only one for which the transition between the time-reversed $+K$ and $-K$ states does not vanish.

In conclusion, given that spin-conserving intervalley transitions consistently explain the recombination of electrons and holes from opposite valleys (see main text), we have attributed the low-energy phonon replica with $K_2$ rather than $K_6$.

**Open discussion**

1. Encapsulation by hBN breaks the mirror symmetry of the system. Thus, further experiments and theory analyses are needed to study if $K_6$ can also contribute to the low-energy phonon replica.
2. First-principle calculation of the matrix elements for spin-conserving intervalley transitions is needed for each of the nine zone-edge phonons. This calculation can be used to evaluate the ratio between the various phonon replicas, and explain the following question: why the phonon replicas $K_1$ and $K_2$ are observed only with the low-energy modes? For example, the selection rule in Eq. (S3) does not discern between the low- and high energy modes of $K_1$ (whose calculated values are 17 and 30 meV). Calculating the matrix elements should shed light on this question.
3. The localization of excitons (and trions) next to defects can enhance their coupling to phonons. Furthermore, defects alleviate the crystal symmetry and enable scattering of exciton complexes with phonon modes other than $K_3$ and $K_1$. As such, localization can explain the observation of the low-energy phonon replica that we have associated to the phonon mode $K_2$. Further experiments and theory analyses are needed to study the effect of localization on the phonon replica.

**SI-7. Magneto PL spectrum with electron doping.**

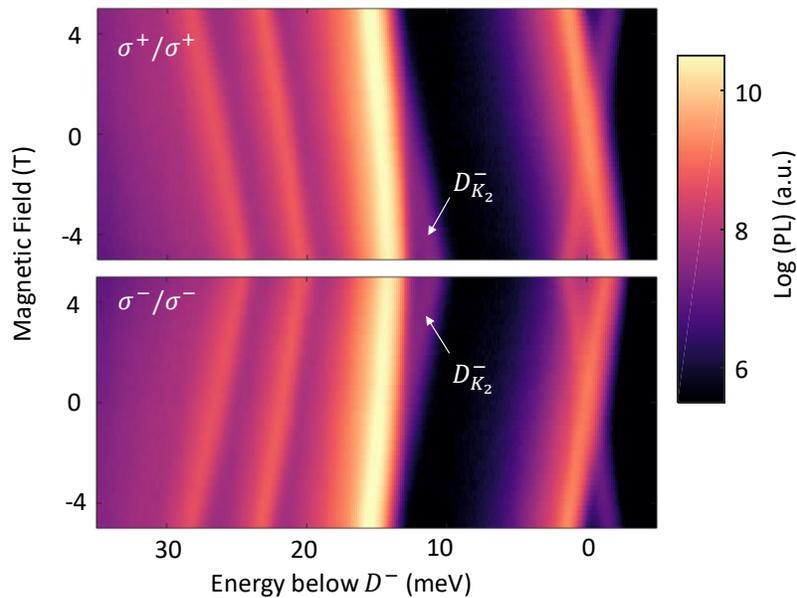

**Figure S7 | Magneto PL spectrum of the triplet emission pattern.** Circular polarization resolved magneto PL spectrum of the triplet states with $\sigma^+/\sigma^+$ (up) and $\sigma^-/\sigma^-$ (down) polarized excitation and detection. The photon energy of the spectrum is offset with respect to $D^-$ at zero field.

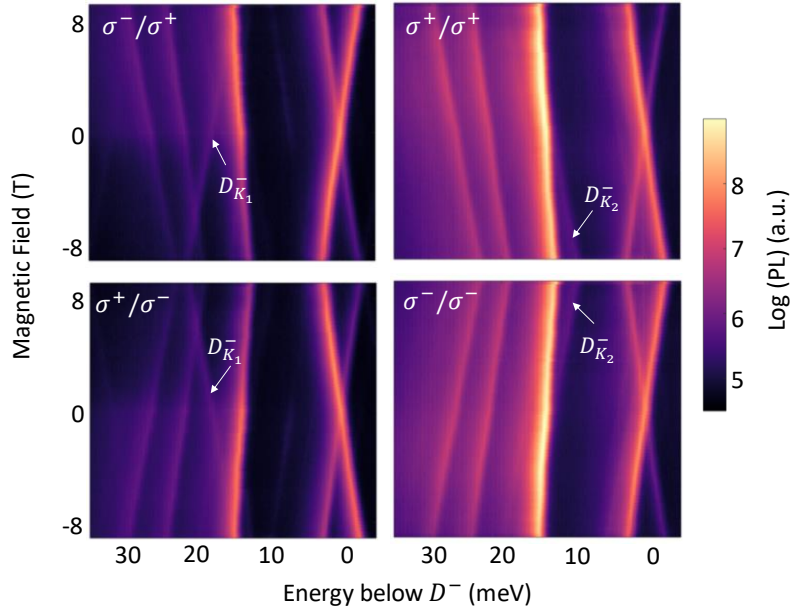

**Figure S8 | Magneto PL spectrum under resonant excitation of the bright neutral exciton.** The sample is electron doped. The polarization configuration of excitation/collection is $\sigma^-/\sigma^+$ (top left), $\sigma^+/\sigma^-$ (bottom left), $\sigma^+/\sigma^+$ (top right), $\sigma^-/\sigma^-$ (bottom right). $D^-_{K_1}$ stands out in the cross polarized spectrum (left panels) with opposite Zeeman energy shift vs B compared with other valley phonon replicas. The faint peak on the higher energy shoulder of $T_1$ at high field corresponds to $D^-_{K_2}$.

## SI-8. Fine structure of the $D^0_{\Gamma_5}$.

Short-range exchange interaction between the electron and hole is predicted to lift the double degeneracy of $D^0$, the dark exciton[12]. Symmetry analysis further shows that the lower energy branch is strictly forbidden whereas the higher energy branch has an out of plane dipole and can give in-plane emission. As shown in Fig.4b, $D^0$ exhibits fine structure, displaying a finite zero field energy splitting, consistent with the previous report.[7]

Remarkably, we also observe the fine structure of the $\Gamma_5$ phonon replica of dark exciton, $D^0_{\Gamma_5}$. In the cross-polarized PL spectrum in Fig. S9a, $D^0_{\Gamma_5}$ shows an energy splitting of 0.6 meV. Magneto-PL spectroscopy further confirms the zero-field energy splitting (Fig. S9c) and reveals the hybridized nature of the two branches at low field. In contrast to the dark exciton $D^0$, which does not show circular polarization, $D^0_{\Gamma_5}$ becomes fully circularly polarized at high field. This shows that whereas the lower energy branch of $D^0$ is strictly forbidden by symmetry, its $\Gamma_5$ phonon replica is allowed due to finite coupling to the bright exciton $X^0$.

This fine feature can also be well captured by the group theory analysis presented in SI-4. The dark (lower energy) and semi-dark (higher energy) branch of dark exciton $D^0$ can be represented by irreducible representation $\Gamma_3$ and $\Gamma_4$ respectively, whereas the bright exciton $X^0$ can be represented by $\Gamma_6$.[11] The selection rule Eq.(S1) then naturally allows the two dark excitons to be coupled to $X^0$ through a zone center $\Gamma_5$ phonon, leading to the observed fine structure of $D^0_{\Gamma_5}$.

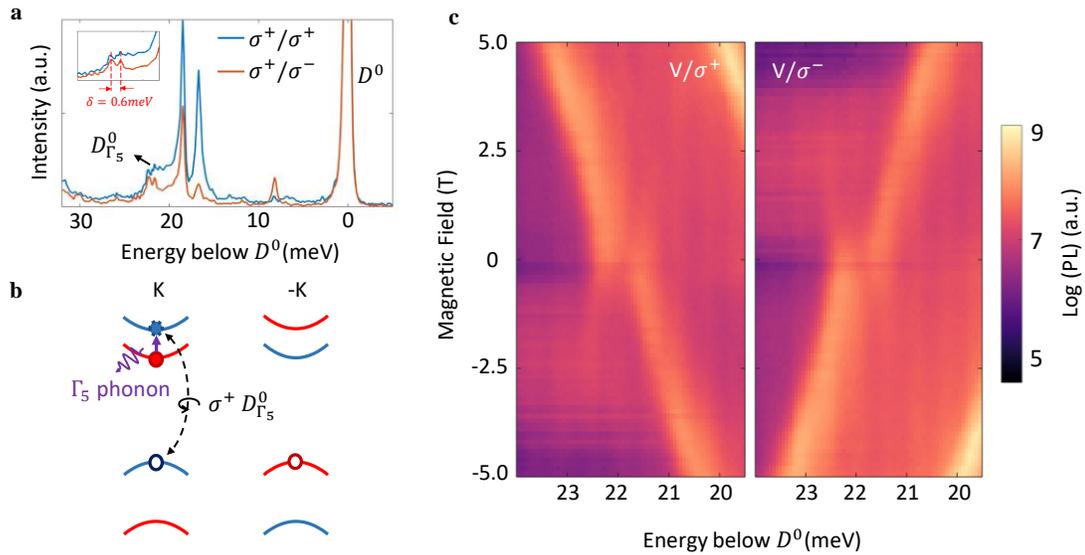

**Figure S9 | Fine structure of the $D^0_{\Gamma_5}$. a,** Circular polarization resolved PL at the charge neutral regime. The photon energy of the spectrum is shifted relative to the dark exciton $D^0$. The cross polarized PL shows the fine structure of $D^0_{\Gamma_5}$, with a splitting $\delta$ of 0.6meV in the peak energy position. **b,** Schematic of the light emission process of a $\sigma^+$ polarized $D^0_{\Gamma_5}$ photon. The electron of the dark exciton $D^0$ first experiences a virtual intravalley spin flip transition while emitting a $\Gamma_5$ phonon. The electron-hole pair then couples with the bright exciton $X^0$, recombines, and emits a $\sigma^+$ polarized photon. **c,** A Zoom in plot of magneto PL spectrum from Fig. 4b, focusing on $D^0_{\Gamma_5}$. The spectrum is taken with linearly polarized excitation, $\sigma^+$ (left) and $\sigma^-$ polarized (right) collection. The photon energy of the spectrum is shifted relative the dark exciton $D^0$.

**Reference**

1. Aivazian, G. *et al.* Magnetic control of valley pseudospin in monolayer WSe$_2$. *Nat. Phys.* **11**, 148 (2015).
2. Srivastava, A. *et al.* Valley Zeeman effect in elementary optical excitations of monolayer WSe$_2$. *Nat. Phys.* **11**, 141 (2015).
3. Li, Y. *et al.* Valley Splitting and Polarization by the Zeeman Effect in Monolayer MoSe$_2$. *Phys. Rev. Lett.* **113**, 266804 (2014).
4. MacNeill, D. *et al.* Breaking of Valley Degeneracy by Magnetic Field in Monolayer MoSe$_2$. *Phys. Rev. Lett.* **114**, 037401 (2015).
5. Xiao, D., Liu, G.-B., Feng, W., Xu, X. & Yao, W. Coupled Spin and Valley Physics in Monolayers of MoS$_2$ and Other Group-VI Dichalcogenides. *Phys. Rev. Lett.* **108**, 196802 (2012).
6. Xu, X. D., Yao, W., Xiao, D. & Heinz, T. F. Spin and pseudospins in layered transition metal dichalcogenides. *Nat. Phys.* **10**, 343-350 (2014).
7. Robert, C. *et al.* Fine structure and lifetime of dark excitons in transition metal dichalcogenide monolayers. *Phys Rev B* **96**, 155423 (2017).
8. Liu, E. *et al.* Chiral-phonon replicas of dark excitonic states in monolayer WSe$_2$. *arXiv preprint arXiv:1906.02323* (2019).
9. Li, Z. *et al.* Emerging photoluminescence from the dark-exciton phonon replica in monolayer WSe$_2$. *Nat. Commun.* **10**, 2469 (2019).

10  Bradley, C. & Cracknell, A. *The Mathematical Theory of Symmetry in Solids: Representation Theory for Point Groups and Space Groups*.  (Oxford University Press, Oxford, 2009).
11  Song, Y. & Dery, H. Transport theory of monolayer transition-metal dichalcogenides through symmetry. *Phys. Rev. Lett.* **111**, 026601 (2013).
12  Dery, H. & Song, Y. Polarization analysis of excitons in monolayer and bilayer transition-metal dichalcogenides. *Phys Rev B* **92**, 125431 (2015).
13  Giannozzi, P. *et al.* QUANTUM ESPRESSO: a modular and open-source software project for quantum simulations of materials. *J. Phys. Condens. Matter* **21**, 395502 (2009).
14  Cardona, M. & Peter, Y. Y. *Fundamentals of Semiconductors*.  (Springer, Berlin, 2005).